\documentclass[aip,jmp,amsmath,amssymb,reprint]{revtex4-1}

\usepackage{graphicx}
\usepackage{dcolumn}
\usepackage{bm}

\usepackage[utf8]{inputenc}
\usepackage[T1]{fontenc}
\usepackage{mathptmx}
\usepackage{amsfonts}
\usepackage{amsthm,latexsym,graphicx,color}

\def\C{\Bbb{C}}
\def\H{\mathcal{H}}
\def\E{\mathcal{E}}
\def\Z{\Bbb{Z}}

\def\T{\Bbb{T}}

\def\R{\Bbb{R}}
\def\G{\Gamma}
\def\H{\mathcal{H}}
\def\pf{{\bf Proof }}
\newtheorem{theorem}{Theorem}[section]
\newtheorem{lemma}[theorem]{Lemma}

\newtheorem{proposition}[theorem]{Proposition}
\newtheorem{definition}[theorem]{Definition}
\newtheorem{remark}[theorem]{Remark}

\begin{document}

\preprint{AIP/123-QED}
\title[Dependence on vertex conditions]{Analyticity of the spectrum and Dirichlet-to-Neumann operator technique for quantum graphs}
\author{Peter Kuchment}
\homepage{http://www.math.tamu.edu/~kuchment}
\affiliation{Mathematics Department, Texas A\&M University, College Station, TX
77843-3368, USA}
\email{kuchment@math.tamu.edu}
\author{Jia Zhao}
\affiliation{Department of Mathematics, School of Science, Hebei University of
Technology, Tianjin, 300401, P.R. China}
\email{zhaojia@hebut.edu.cn}
\thanks{P.K. acknowledges support of NSF DMS-1517938 grant, J.Z acknowledges support of Hebei University of Technology and Texas A\&M University's hospitality during her visit.}
\date{\today}
\begin{abstract}
In some previous works, the analytic structure of the spectrum of a quantum graph operator as a function of the vertex conditions and other parameters of the graph was established. However, a specific local coordinate chart on the Grassmanian of all possible vertex conditions was used, thus creating an erroneous impression that something ``wrong'' can happen at the boundaries of the chart. Here we show that the analyticity of the corresponding ``dispersion relation'' holds over the whole Grassmannian, as well as over other parameter spaces.

We also address the Dirichlet-to-Neumann (DtN) technique of relating quantum and discrete graph operators, which allows one to transfer some results from the discrete to the quantum graph case, but which has issues at the Dirichlet spectrum. We conclude that this difficulty, as in the first part of the paper, stems from the use of specific coordinates in a Grassmannian and show how to avoid it to extend some of the consequent results to the general situation.
\end{abstract}
\maketitle
\section*{Introduction}
Analytic structure of the spectrum of a quantum graph operator as a function of the vertex conditions and other parameters of the graph was studied in \cite{BerKuc_arx10,BKbook}. However, a specific form of vertex conditions was used, which corresponds to a local coordinate chart on a related Grassmannian. This created an erroneous impression that something ``wrong'' can happen when some parameters of the vertex conditions go to zero or infinity (i.e., at the boundaries of the chart). Here we show that the analyticity of the corresponding ``dispersion relation'' holds over the whole Grassmannian of vertex conditions, as well as over other parameter spaces.

We also address the Dirichlet-to-Neumann (DtN) technique of relating quantum and discrete graphs operators, which allows one to transfer some results from the discrete to the quantum graph case, but which has issues at the Dirichlet spectrum. Although this might seem to be different from the above consideration, we conclude that this difficulty also stems from the use of specific coordinates in a Grassmannian and show how to avoid it to extend some of the consequent results to the general situation.

The paper is structured as follows: Section \ref{S:basic} covers relevant basic information about quantum graphs. In section \ref{S:grass} we explain the relation between Grassmannians and vertex conditions, consider the relevant analytic bundles over the Grassmannians, and establish as one of our main results the analytical dependence on vertex conditions result in Theorem \ref{T:analytic}. The proof of the theorem is given in section \ref{S:proof}. In section \ref{S:DtN} we turn to the second problem of the paper - reducing quantum graph problems to discrete ones. here we sketch the standard DtN procedure and provide the crucial (well know) Lemma \ref{L:equiv}. Then a modified DtN procedure, which can be used for any quantum graph, is described, resulting in Theorem \ref{T:modDtN}. In the next section, this leads to establishing the ``cleaned up'' from unnecessary conditions very useful Theorem \ref{T:DtN} on the structure of bound states of periodic quantum graphs. One should also look at the remark in section \ref{S:GrAgain}, where a different from the modified DtN above general construction is described (kudos to G.~Berkolaiko). Section \ref{S:remarks} contains some extended remarks and generalizations.
\section{Basic notions}\label{S:basic}

We recall here some basic notions concerning quantum graphs, see e.g. \cite{BKbook}. The graph $\G$ has finitely many vertices $v\in V$ and edges $e\in E$. We use the notations $|V|$ and $|E|$ for the number of vertices and edges correspondingly. Each edge $e$ is endowed with a coordinate $x_e$ that identifies $e$ with a finite segment $[0,l_j]\subset \R$. This allows one to consider differential expressions along the edges, such as for instance $\dfrac{d}{dx_e}$ on sufficiently smooth functions on $\G$. We will concentrate in particular on the second order ``Laplacian'' $L:=-\dfrac{d^2}{dx_e^2}$, albeit much more general operators of second and higher orders can be considered without a problem, see Section \ref{S:remarks}. The functions in the domain of the operator are assumed to belong to the space
\begin{equation}
\widetilde{H^2}(\G):=\bigoplus\limits_{e\in E} H^2(e).
\end{equation}
Additional conditions on the values of a function and its first derivatives are usually enforced at the vertices. For the reasons to be explained later, the total number of these conditions is assumed to be equal to $2|E|$. In particular, in so-called ``local'' conditions, the number of conditions imposed at a vertex is equal to the degree of the vertex and they involve only values at the vertex.

The ``dispersion relation,'' i.e. the graph of the spectrum $\sigma(L)$ of the operator as a multiple-valued function of the vertex conditions and edges' lengths was shown in \cite{BerKuc_arx10} (see also \cite[Section 2.5]{BKbook}) to be an analytic variety. However, a smaller subset of vertex conditions was excluded there, which was an artifact of the proof. We show here (Theorem \ref{T:analytic}) that such analyticity holds over the whole Grassmannian of vertex conditions, as well as over the spaces of potentials and edges' lengths.

In order for the spectrum to have a useful meaning, it is necessary that the Fredholm index of the resulting operator $L$ is equal to zero. Although Fredholm property is automatic in our situation \cite{FulKucWil_jpa07,BKbook}, the index might be different from zero. As shown in \cite{FulKucWil_jpa07} (see also \cite[Theorem 2.4.3]{BKbook}), to assure that the index is indeed equal to zero requires that the total number of vertex conditions is equal to $2|E|$, which explains why we have imposed this restriction.

\begin{figure}
  \centering
  \includegraphics[scale=0.5]{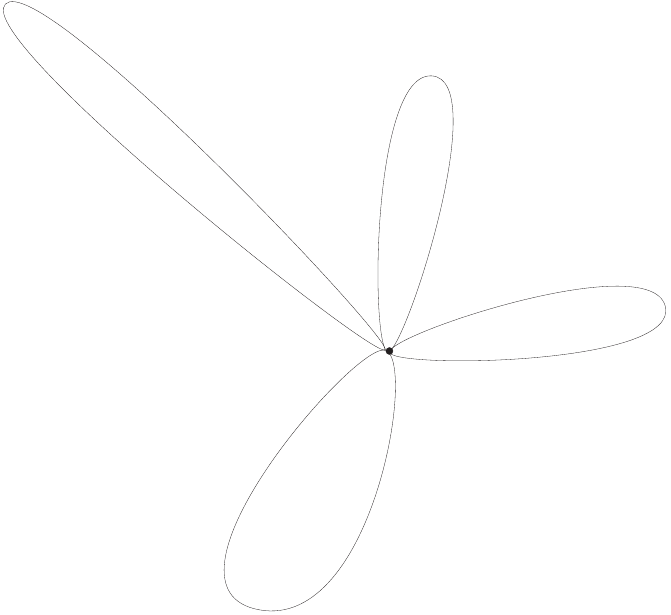}
  \caption{A rose}\label{F:rose}
\end{figure}

Another comment is that, as it is well known (and shown, for instance, in \cite[Section 1.4.6]{BKbook}), any finite quantum graph can be ``folded'' into a ``rose'' with a single vertex and $|E|$ loop ``petals.'' This would cover both cases of local and non-local vertex conditions. In particular, proving our results for such a rose immediately implies it for local conditions as well. We thus concentrate for now on a rose $\G$ with a single vertex $v$, $|E|$ petals, and $2|E|$ conditions imposed at $v$ (see Fig. \ref{F:rose}).

%


\section{Grassmannians and vertex conditions}\label{S:grass}

Consider the rose $\G$ (Fig. \ref{F:rose}) with a single vertex $v$, $|E|$ petals, and $2|E|$ conditions imposed at $v$.

The space of all values of functions $f\in \widetilde{H^2}(\G)$ and their first derivatives at the vertex $v$ can be identified with $\C^{4|E|}$. Imposing $2|E|$ conditions at $v$ selects a $2|E|$-dimensional subspace $P\subset \C^{4|E|}$. We will identify this subspace with the set of vertex conditions, and thus it defines a closed subspace
\begin{equation}
H(P)\subset \widetilde{H^2}(\G)
\end{equation}
consisting of functions with the vertex values belonging to the subspace $P$,
and the operator
\begin{definition}
\begin{equation}
L(P):=L|_{H(P)}.
\end{equation}
\end{definition}
According to \cite[Theorem 2.4.3]{BKbook}, this operator acting from $H(P)$ to $L_2(\G)$ is Fredholm and of index zero and thus might have a non-trivial spectral theory.

The subspace $P\subset \C^{4|E|}$ is a point of the complex Grassmannian $G(2|E|,4|E|)$, which is well known to be a compact analytic (projective) variety (see, e.g., \cite[Ch. 3]{Weyman}). Thus, our main goal is to understand the following object:

\begin{definition}\underline{For the purpose of this text}, the \textbf{dispersion relation} of the operator $L$ is defined as follows:
\begin{equation}
\begin{split}
D_L:=\{(P,\lambda)\in G(2|E|,4|E|)\times \C\, |\, \mbox{ operator }\\
(L(P)-\lambda I) \mbox{ does not have bounded inverse}\}.
\end{split}
\end{equation}
\end{definition}

\begin{remark} Another name for $D_L$ is the \textbf{spectrum of the Fredholm operator-function} (see, e.g., \cite{ZaiKreKucPan_umn75}.)
\begin{equation}
(P,\lambda)\in G\times \C \mapsto(L(P)-\lambda I),
\end{equation}
where we use the shorthand notation $G:=G(2|E|,4|E|)$.
\end{remark}

Our main result is the following
\begin{theorem}\label{T:analytic}
The dispersion relation $D_L$ is a principal analytic subset of $ G(2|E|,4|E|)\times \C$.

I.e., it can be locally described by equations of the form $f(P,\lambda)=0$ with scalar analytic functions $f$.
\end{theorem}

The proof of this result is postponed till section \ref{S:proof}.

So far, self-adjointness condition has not been enforced on the operator $L(P)$. It has been known (and rediscovered) for a long time, though (see  \cite{EveMar_mams01,EveMar_mams04,Everitt_bvp,Har_jpa00,Naimark_ldo,Nov_incol99,AkhiezerGlazman_linop}and the references in \cite[Appendix C]{BKbook}) that self-adjointness is equivalent to the subspace $P$ being Lagrangian with respect to the natural complex symplectic structure on $\C^{4|E|}$. The corresponding Lagrangian Grassmannian is a (real) analytic sub-manifold in $ G(2|E|,4|E|)$, so Theorem \ref{T:analytic} applies by restriction.

We consider now some vector bundles over the Grassmannian $G:=G(2|E|,4|E|)$.

\begin{definition}\label{D:tautology}
The \textbf{tautological bundle} $\T\mapsto G$ over $G$ consists of points $(P,u)\in G\times \C^{4|E|}$ such that $u\in P$.
\end{definition}
The following statement is well known and can be proven by constructing in local Pl\"ucker coordinates in $G$ \cite{Weyman} a projector $Q(P):\C^{4|E|}\mapsto P$ that depends analytically on $P\in G$. For readers' convenience, we present its brief proof.

\begin{lemma}\label{L:analytic}
$\T$ is an analytic sub-bundle of the trivial bundle $G\times \C^{4|E|}$ over $G$.
\end{lemma}
\textbf{Proof}. Let $P_0\in G(k,n)$, with $n>k$. Then one can consider $P_0$ as a $k$-dimensional subspace in
$\C^n$: $P\subset \C^n$.

Let $R\subset \C^n$ be any ($n-k$)- dimensional subspace in $\C^n$ complementary to $P_0$, thus
$$
\C^n=P_0\bigoplus R.
$$
In this direct decomposition, one can consider $P_0$ as the graph of the zero linear operator from $P_0$ to $R$. Then $k$-dimensional subspaces $P\in G(k,n)$ close to $P_0$ in Grassmannian's topology are in one-to-one correspondence with graphs of linear operators $A:P_0\mapsto R$ of small norm. The matrix elements of $A$ are local coordinates near $P_0\in G(k,n)$.

Now one considers the following linear operator:
$$Q(P): (x,y)\in P_0\bigoplus R \mapsto (x,Ax).$$
It is clearly a projector on the subspace $P\in\C^n$, which depends analytically on $A$.

This finishes the proof of the Lemma. $\Box$
Consider now the trivial Hilbert bundle
$$\pi: G\times \widetilde{H^2}(\G)\mapsto G$$
and its subset
$$\H:=\{(P,u)\in G\times \widetilde{H^2}(\G)\, |\, u\in H(P) \}.$$

\begin{proposition}\label{P:anal}
$\H$ is an analytic Hilbert bundle over $G$ with respect to the projection $\pi$.
\end{proposition}
\pf The proof will follow (see, e.g., \cite{ZaiKreKucPan_umn75}), if one constructs a bounded projector
$$R(P): \widetilde{H^2}(\G)\mapsto H(P)$$
that depends (locally) analytically on $P\in G$.

Let us denote by $S:\widetilde{H^2}(\G) \mapsto \C^{4|E|}$ the operator of taking the traces of the function and its first derivative at the sole vertex $v$ from all incoming edge directions. We also denote by $\E$ an operator that extends the vertex values from $\C^{4|E|}$ by linearity along the edges (for a small distance, to avoid conflicts).

Let also $\phi(x)$ be a scalar function on $\G$ that is equal to $1$ in a small neighborhood of $v$, is smooth along the edges, and vanishes outside a somewhat larger neighborhood of $v$.

We now define the operator as follows:
$$(R(P)f)(x):=f(x)-\phi(x)\E(I-Q(P))S(f).$$
It is easy to see that it is a projector onto $H(P)$, which depends on $P$ as well as $Q(P)$ does, i.e. (locally) analytically. $\square$

\section{Proof of Theorem \ref{T:analytic}}\label{S:proof}

The operator $L(P)$ is the restriction of a fixed bounded operator $L:\widetilde{H^2}(\G)\mapsto L_2(\G)$ onto the subspace $H(P)$. According to Proposition \ref{P:anal}, this implies that the resulting mapping from the analytic Hilbert bundle $\H$ to the trivial bundle $G\times L_2(\G)$ is an analytic Fredholm morphism. Hence, the family $L(P)-\lambda I$ is an analytic Fredholm morphism of Hilbert bundles over $G\times \C$. Now the general Theorem 4.11 in \cite{ZaiKreKucPan_umn75} (see also
\cite[Theorem 1.6.16]{Kuc_floquet} and \cite[Theorem B.3.3]{BKbook}) implies the required analyticity and thus finishes the proof (see \cite{ZaiKreKucPan_umn75} for extended discussion of this topic). $\square$



\section{Dirichlet-to-Neumann maps and relations between quantum and discrete graphs.}\label{S:DtN}

In this section we tackle another issue that has been causing problems. Namely, the use of the Dirichlet-to-Neumann map to reduce quantum graph problems to the corresponding discrete ones.

\subsection{Relations between quantum and discrete graph spectra and Dirichlet-to-Neumann map}

We recall the notion of the Dirichlet-to-Neumann map (see, e.g. \cite{Cat_mm97} and \cite[Section 3.5]{BKbook}).

Let $\Gamma$ be a finite quantum graph (or a $\Z^n$-periodic one with a compact fundamental domain). Then the number of different edge lengths is finite. Consider for each edge the (discrete, bound from below) spectrum of $-d^2/dx^2$ with Dirichlet conditions at the ends. The union of these spectra we denote by $\sigma_D(\Gamma)$.

If $\lambda\notin \sigma_D$, then one can define on each edge $e$ the Dirichlet-to-Neumann map $D(\lambda)$ that starts from the give vertex values $(u(0),u(l_e))$, finds the solution of the problem $-u''(x)-\lambda u(x)=0$ on the edge, and then ends up with the Neumann data $(-u'(0),u'(l_e))$. As the result, the set $U'$ of all vertex values of derivatives of $u$ is expressed as $D(\lambda)U$, where $U$ is the set of all vertex values of $u$.

Now, vertex conditions (in the form provided by Kostrykin and Schrader, see \cite{BKbook} can be written as follows:
$$AU+BU'=0$$
for appropriate matrices $A$ and $B$. Hence, this can now be written as
\begin{equation}\label{E:discr}
AU+BD(\lambda)U=0.
\end{equation}
In other words our spectral problem for the original graph is reduced to the \underline{equivalent} discrete (while non-linearly dependent upon $\lambda$) graph problem (\ref{E:discr}).

The following statement is straightforward:
\begin{lemma}\label{L:equiv}
If $\lambda\notin\sigma_D$, then
\begin{enumerate}
\item the DtN correspondence between the solutions of the quantum graph and discrete eigenvalue problems is a bijection;
\item if the graph is periodic, under this bijection $L_2$-solutions on the quantum graph correspond to $l_2$-solutions on the discrete one;
\item under this bijection compactly supported solutions correspond to compactly supported solutions.
\end{enumerate}
\end{lemma}

This procedure does not work when $\lambda$ is in the spectrum of Dirichlet Laplacian on an edge.

\subsection{A modified DtN procedure}

Let us introduce the following modified construction, involving adding some new ``fake'' vertices.

Consider an eigenvalue problem for a quantum graph $Lu=\lambda u$, where $L$ is the second order operator $-\dfrac{d^2}{dx^2} +V(x)$ with a potential $V$ that belongs to $L_\infty$ on each edge. Let us fix a value $\lambda$, which is allowed to belong to $\sigma_D(\Gamma)$ and consider an edge $e$. If $\lambda$ is not a Dirichlet eigenvalue on this particular edge $e$, then no action takes place. If it is, though, we insert a new vertex (a \textbf{dot}) $x_e$ into $e$ in such a way that $\lambda$ is not a Dirichlet eigenvalue of either of the two resulting sub-segments. This is always possible, as long as one avoids the (finitely many) zeros of the corresponding eigenfunction. We then impose the Kirchhoff condition at $x_e$. This means that the solutions to the left and right of $x_e$ match with their first derivatives. In other words, in the quantum graph case these vertices can be ignored. We, however, want to keep them. The new graph (with one extra vertex in each edge where $\lambda$ is a Dirichlet eigenvalue) we will call the \textbf{dotted graph} $\dot{\Gamma}$.
\begin{figure}[ht!]
  \centering
  \includegraphics[scale=0.8]{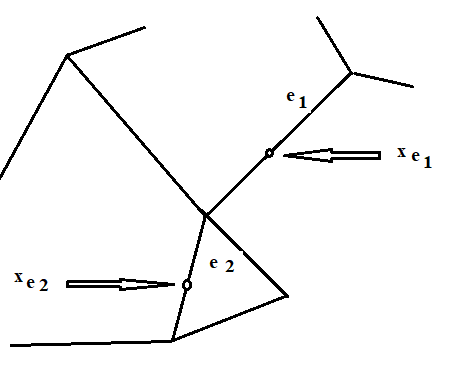}
  \caption{A graph with extra vertices (dots) inserted into two edges.}\label{F:dot}
\end{figure}

Let us now assume the graph to either being finite, or periodic (i.e., having the group $\Z^n$ acting on it freely and co-compactly). For any such graph and any $\lambda$, the choices of the extra vertices (dots) can be made in such a way that $\lambda\notin \sigma_D(\dot{\Gamma})$. Notice that in the periodic case the dotted graph can be assumed to be periodic as well. Indeed, the dots can be chosen in the fundamental domain only and then distributed periodically.
Now the standard DtN procedure works for the dotted graph $\dot{\Gamma}$, since one can deduce the following result from Lemma \ref{L:equiv}:
\begin{theorem}\label{T:modDtN}
\begin{enumerate}
\item The modified DtN correspondence between the solutions of the quantum graph and discrete eigenvalue problems is a bijection;
\item if the graph is periodic, under this bijection $L_2$-solutions on the quantum graph correspond to $l_2$-solutions on the discrete one;
\item under this bijection compactly supported solutions correspond to compactly supported solutions.
\end{enumerate}
\end{theorem}

\subsection{Bound states on periodic quantum graphs}
This problem with using the DtN map has arisen, for instance, when studying spectra of graphene and carbon nanotubes in \cite{KucPos_cmp07}. In particular, the following result for discrete periodic graphs
\cite{Kuc_incol89} (see also \cite[Theorem 4.5.2]{BKbook}) holds:
\begin{theorem}\cite{Kuc_incol89}\label{T:olddiscr}
If a periodic discrete equation $Au=0$ on a periodic graph $\Gamma$ has a non-zero solution in $l_2(\Gamma)$, then it also has a non-zero compactly supported solution. Moreover, the compactly supported solutions form a complete set in the space of all $l_2$-solutions.
\end{theorem}
However, while using the Dirichlet-to-Neumann map to transfer it to the quantum graph case, one arrives only to the following
\begin{theorem}\cite{Kuc_jpa05} (see also \cite[Theorem 4.5.4]{BKbook}
Let $\Gamma$ be a $\Z^n$-periodic quantum graph equipped with the second derivative Hamiltonian and arbitrary vertex conditions at the vertices. Suppose that $\lambda$ is not a Dirichlet eigenvalue on any edge. Then, existence of a non-zero $L_2$ eigenfunction corresponding to the eigenvalue $\lambda$ implies existence of a compactly supported eigenfunction, and the set of compactly supported eigenfunctions is complete in the (infinite-dimensional) eigenspace.
\end{theorem}
Here the restriction of avoiding Dirichlet eigenvalues has appeared due to the use of the DtN technique (DtN operator is not defined on the Dirichlet spectrum). However, in many cases Dirichlet eigenvalues are exactly of interest (e.g., \cite{KucPos_cmp07,Ngoc}).
So, the question arises whether this restriction is an artifact of the proof.

Indeed, it does happen to be, as the following result shows:
\begin{theorem}\label{T:DtN}
Let $\Gamma$ be a $\Z^n$-periodic quantum graph equipped with the second derivative Hamiltonian and arbitrary vertex conditions at the vertices. Existence of a non-zero $L_2$ eigenfunction corresponding to an eigenvalue $\lambda$ implies existence of a compactly supported eigenfunction, and the set of compactly supported eigenfunctions is complete in the (infinite-dimensional) eigenspace.
\end{theorem}

The proof is rather straightforward. Indeed, we only need to handle the case when $\lambda\in\sigma_D$. In this case, we consider the dotted periodic quantum graph $\dot{\Gamma}$, where, as we have mentioned before, the spectral problem is equivalent to the one on the original graph. Using the DtN procedure, we arrive to a discrete spectral problem on the periodic discrete graph $\dot{\Gamma}$, where the Theorem \ref{T:olddiscr} applies. Now Lemma \ref{L:equiv} finishes the proof. $\Box$

\section{Final remarks and conclusions}\label{S:remarks}
\subsection{Local vertex conditions}
Most works on quantum graphs assume only local vertex conditions, i.e. those that relate only the values of the function and its edge derivatives at one vertex at the time. Thus, vertex conditions at each vertex correspond to the points of the ``local'' Grassmannian $G_v$, and the totality of all local vertex conditions corresponds to the points of the product space $\prod\limits_{v\in V}G_v$, which is an analytic submanifold of the Grassmannian $G$ of all (non-local) conditions considered in Section \ref{S:grass}. Thus, by restriction, Theorem \ref{T:analytic} applies under the restriction of local vertex conditions as well.

\subsection{Self-adjoint case}
The similar conclusion on applicability of Theorem \ref{T:analytic} as above holds under the requirement of self-adjointness of the operators, since the corresponding Lagrangian Grassmannian of vertex conditions is a real-analytic submanifold in $G$

\subsection{Dependence on potentials and edge lengths}
Besides the vertex conditions $P\in G$, as before, the spectrum of a quantum graph operator $L$ depends on other parameters, for instance the potential $V$ in $L=-\dfrac{d^2}{dx^2}+V(x)$ and edge lengths $\{l_e\}_{e\in E}\in (\R^+)^{|E|}$. In fact, sacrificing self-adjointness, one can allow the edge lengths to be any non-zero complex numbers (see \cite[Section 3.1.2]{BKbook}), so then $\xi:=\{l_e\}_{e\in E}\in (\C^\setminus \{0\})^{|E|}$.
Thus, we denote the corresponding operator by $L(P,V,\xi)$, where
$$
(P,V,\xi)\in G(2|E|,4|E|)\times L_\infty(\Gamma)\times (\C \setminus \{0\})^{|E|}).
$$
Correspondingly, the (extended) dispersion relation is defined as follows:
\begin{equation}
\begin{split}
D_L:=\{(P,V,\xi,\lambda)\in G(2|E|,4|E|)\times L_\infty(\Gamma)\times (\C \setminus \{0\})^{|E|})\times \C\,| \\
\mbox{ such that } (L(P,V,\xi)-\lambda I) \mbox{ does not have a bounded inverse}\}
\end{split}
\end{equation}
Then the analog of Theorem \ref{T:analytic} holds:
\begin{theorem}
The dispersion relation $D_L$ is a principal analytic subset in $G(2|E|,4|E|)\times L_\infty(\Gamma)\times (\C \setminus \{0\})^{|E|})\times \C$
\end{theorem}
The proof of Theorem \ref{T:analytic} adjusts easily to the presence of these extra parameters.

Clearly, the above comments about local vs. non-local conditions and self-adjointness still apply.
\subsection{Higher order operators}

The considerations and results of Sections \ref{S:grass}-- \ref{S:proof} transplant without any change to higher order differential operators on metric graphs, such as for instance in \cite{FulKucWil_jpa07} and \cite[Section 2.4]{BKbook}. One needs to assume the number of vertex conditions to be right, to guarantee the index of the operator to be equal to zero (\cite{FulKucWil_jpa07} and \cite[Section 2.4]{BKbook}). Otherwise, the analyticity result holds vacuously: the spectra would fill the whole complex plain.

\subsection{Grassmannian again?}\label{S:GrAgain}
It might look like the considerations of Section \ref{S:DtN} are not much related to the ones in the first part of the paper. One can argue however, that (although it is not immediately obvious) like in the first part of this text, the culprit is a specific choice of local coordinates on a Grassmannian.
Indeed, what the DtN map does, it removes the ODE on each edge, replacing it with a relation between the data at the vertices. This, however, can be done in very general terms, without involving solutions of the Dirichlet boundary value problem.

Namely, suppose we are interested in solutions of a second order differential equation $Lu=0$ on the edges, with vertex conditions represented by a plane $P\in G(2|E|,4|E|)$, as in Section \ref{S:grass}.   Let us look at an edge $e$ with vertices $v_1,v_2$ and consider the $4$-dimensional space $B_e$ of vertex values $(f(v_1),f'(v_1),f(v_2),f'(v_2))$ of functions  $f\in H^2(e)$, where the derivatives are taken in the directions into the edge. There is a natural surjection $\pi: H^2(e) \mapsto B_e$. Let the two-dimensional subspace $S_e\in H^2(e)$ consist of all solutions of $Lu=0$ on $e$. Consider its image under $\pi$: $P_e:=\pi(S_e)$. It is clear that $\pi$ is injective on $S_e$, and thus $P_e$ is two-dimensional.

We now denote by $Q$ the subspace in $\C^{4|E|}$ consisting of vectors, whose restrictions to $v_1,v_2$ belong to $P_e$ for any edge $e$ connecting the vertices $v_1$ and $v_2$. In other words, satisfying equation $Lu=0$ on the edge $e$ is equivalent to the inclusion $\pi(u)\in Q$ and satisfying the vertex conditions, as we know, is equivalent to being in $P$. This leads to the following simple result:
\begin{theorem}
A function $f$ satisfies the equation $Lf=0$ with the vertex conditions corresponding to the subspace $P$ if and only if the vector of its vertex values
$$\phi:=\{(f(v_1),f'(v_1),f(v_2),f'(v_2))\}_{e=(v_1,v_2)\in E}\in \C^{4|E|}$$
belongs to the subspace
\begin{equation}\label{E:discr2}
\Phi:=P\bigcap Q.
\end{equation}
\end{theorem}
Since the inclusion $u\in\Phi$ can be written as a system of linear equations for $u$
\begin{equation}\label{E:discr3}
A(u)=0,
\end{equation}
the equation $Lf=0$ with the vertex conditions corresponding to the subspace $P$ is converted into a system of discrete equations.

More specifically, suppose one chooses on each edge a fundamental set
of solutions, ${f_{e,1}, f_{e,2}}$ and let $f$ be their superposition  $f = a_{e,1}
f_{e,1} + a_{e,2} f_{e,2}$. Then inclusion into $Q$ is automatically guaranteed, so substituting $f$ into the vertex conditions $P$ leads to a discrete problem for the set of coefficients $\{a_{e,1},a_{e,2}\}_{e\in E}$.

For instance, if the operator is $L=-\dfrac{d^2}{dx^2}-k^2$, one can choose as the fundamental solutions functions $f_1: = \sin(k (l_e-x)) / \sin(k l_e), f_2 := \sin(k x) / sin(k l_e)$. One can check that this leads to the standard DtN procedure.

If one chooses  $f_1: = \exp(ikx),
f_2 = \exp(ik(l_e-x))$, one arrives at the bond scattering matrix description of a quantum graph \cite{KotSmi_ap99}.

For this procedure, an analog of Lemma \ref{L:equiv} also holds, so in particular it can be used to provide an alternative proof of Theorem \ref{T:DtN}.

\subsection{Some other references}
Besides the sources referred to before, considerations somewhat related to the ones of Section \ref{S:grass} can be found in \cite{ZhaoJia_bc,Kong1,Kong2,Kong3}

\section{Acknowledgments}
%
The first author expresses his gratitude to the NSF for the grant support. The second author thanks Hebei University of Technology for the financial support and Texas A\&M University for hospitality during a visit. Both authors thank G.~Berkolaiko, who independently came up with the approach of Section \ref{S:GrAgain}, for inspiring discussions. They also express their gratitude to J.~M.Landsberg, F.~Sottile, and M.~Zaidenberg for useful comments and references.

\bibliographystyle{abbrv}
\bibliography{bk_bibl}

\end{document}